\documentclass[twocolumn,twoside,slac_two]{revtex4}
\usepackage{graphicx}
\usepackage{fancyhdr}
\newcommand{\hess}{\textsc{H.E.S.S.}}

\newcommand{\fer}{{\sl {\it Fermi}}}
\newcommand{\fla}{\fer-LAT}

\newcommand\arcsec{\mbox{$^{\prime\prime}$}}%
\begin{document}

%Title of paper
\title{Multiwavelength campaign on the HBL PKS~2155-304 :\\ A new insight on its spectral energy distribution}

% Repeat the \author .. \affiliation  etc. as needed
%
% \affiliation command applies to all authors since the last
% \affiliation command. The \affiliation command should follow the
% other information

\author{D.~A.~Sanchez}
\affiliation{Laboratoire d'Annecy-le-Vieux de Physique des Particules, Universit\'e de Savoie, CNRS/IN2P3}
\author{B.~Giebels, D.~Zaborov}
\affiliation{Laboratoire Leprince-Ringuet/\'Ecole Polytechnique/IN2P3/CNRS}

\author{R.~D.~Parsons}
\affiliation{Max-Planck-Institut fur Kernphysik}

\author{G.~M.~Madejski, A.~Furniss}
\affiliation{Kavli Institute for Particle Astrophysics and Cosmology,Department of Physics and SLAC National Accelerator Laboratory}

\author{On behalf of the NuSTAR, \fer\ and \hess\ collaborations}

\begin{abstract}

The blazar PKS~2155-304 was the target of a multiwavelength campaign from June to October 2013 which widely improves our knowledge of its spectral energy distribution. This campaign involved the NuSTAR satellite (3-79 keV), the \textit{Fermi} Large Area Telescope (LAT, 100~MeV-300~GeV) and the High Energy Stereoscopic System (H.E.S.S.) array phase II (with an energy threshold of few tens of GeV). While the observations with NuSTAR extend the X-ray spectrum to higher energies than before, H.E.S.S. phase II, together with the use of the LAT  {\tt PASS 8}, enhance the coverage of the $\gamma$-ray regime with an unprecedented precision. In this work, preliminary results from the multi-wavelength analysis are presented. 
\end{abstract}

%\maketitle must follow title, authors, abstract
\maketitle

\thispagestyle{fancy}

\section{Introduction}
Several new and upgraded instruments have come online in the last few years. In
the X-ray regime, NuSTAR, using true focusing optics, is providing an unprecedented
view of the hard-X-ray sky with its wide energy range (3-79 keV). In the
very-high-energy $\gamma$-ray regime, H.E.S.S. is now operating with five
telescopes including the largest Cherenkov telescope ever constructed.
Additionally, a major upgrade to the \fla\ instrumental response functions
(IRFs), {\tt PASS 8} \cite{p8}, is now being implemented, increasing the
sensitivity on the high-energy $\gamma$-ray sky.

The high frequency peaked BL Lac (HBL) object PKS 2155-304 was
the target of a new multi-wavelength campaign from April to October 2013 involving these three
instruments. This campaign provides a more complete coverage of the X-ray and
$\gamma$-ray range than the previous campaign held in 2008, which involved \fer,
H.E.S.S. and also ATOM, \textit{Swift} and \textit{RXTE} and lasted for 11 days \cite{2008}.

NuSTAR observed PKS~2155-304 multiple times, starting with a $\approx$40\,ks
observation in April 2013 designed for cross-calibration purposes of various
high energy astrophysical instruments. Seven subsequent observations (lasting
$\approx$10\,ks each), in July, August, and September 2013, were scheduled to be
strictly simultaneous with \hess, during local moonless night-time periods at
the \hess\ location. Furthermore, some independent observations of PKS~2155-304
were conducted with \hess\ for calibration and monitoring purposes. In its
normal operation mode, the \fla\ is observing the full sky and each source is
seen 30 minutes every 3 hours.

\section{Observations}
\subsection{\hess\ Mono-mode}

The H.E.S.S. experiment consists of an array of 5 telescopes: four $12m$
diameter dish telescopes in operation since 2004 and and a fifth, CT5, a $28m$
diameter dish, in operation since 2012. This yields an energy threshold of the
instrument to be of the order of a few tens of GeV. The results of the first observations
conducted with the fifth H.E.S.S. telescope standalone (in the so-called
Mono-mode) are here presented in this work.

The data have been analysed with the {\tt Model} analysis \cite{model} with cuts
adapted for this telescope. PKS~2155-304 is detected at a level of 26.4$\sigma$
in 35 hours of live time. The spectrum is well described by a simple power-law
with an energy threshold of 98~GeV: 
$$F(E) =(26.1\pm 1.2)\times 10^{-10} {\rm cm}^{-2}\,{\rm s}^{-1} E_*^{-(3.25\pm0.08)} $$
with $E_*=E/(196\ {\rm GeV})$ the decorrelation energy. This spectrum
is consistent with the source being at a low flux state, lower than the flux measured
in 2008. Figure \ref{SED} presents the spectral energy distribution (SED) measured with CT5.

Variability has been found on the night-by-night light curve with a fractional excess variance $F_{\rm var} = 67\pm10\%$, higher than what has been found in 2008.

\subsection{\fla\ data}

The \fla\ \cite{fermi} data have been analysed using the ScienceTools {\tt v9R34P1} and the new {\tt PASS 8} IRFs yielding an extended energy range with respect to previous analysis and an increased effective area, better PSF and background rejection with respect to the {\tt PASS 7} IRFs. 

\begin{figure}
\includegraphics[width=85mm]{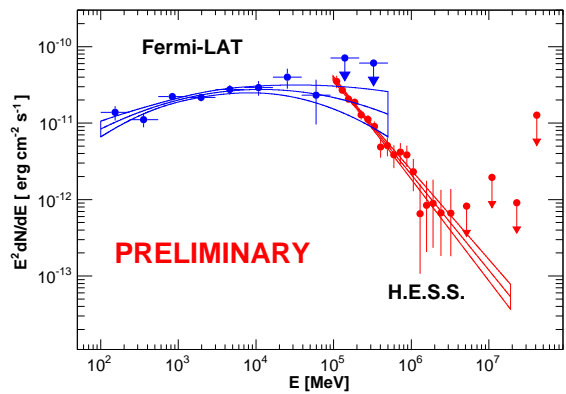}
\caption{Spectral energy distribution of PKS~2155-304 obtained with \fla\ in blue and H.E.S.S. in red. Contours give the 1 $\sigma$ error band.}
\label{SED}
\end{figure}

Photons with an energy from 100~MeV to 500~GeV region of interest within 15 degrees of the source
 coordinates were used. Data have been analysed using a binned  maximum likelihood analysis implemented in {\tt gtlike}. The sky model has been built using the 4 years catalogue of point sources \cite{3FGL} and the spectrum is found to be best fitted by a log-parabola model of the form:
$$F(E) \propto E_*^{-(1.73\pm0.07) -(0.05\pm0.02)~log_{10}E_*}$$
where $E_*=E/(904\ {\rm MeV})$, for an integrated flux above 100~MeV $I=(8.15\pm0.89)\times 10^{-8}$cm$^{-2}$\,s$^{-1}$ (see Fig. \ref{SED}). 

The spectrum is found to be consistent with results obtained with {\tt PASS 7} IRFs but with increased statistic. No variability has been found in the \fer\ light curve, nevertheless the flux is lower than the one measured in 2008 during the first \fer-\hess\ campaign on this object.

\subsection{NuSTAR observations and results}

NuSTAR \cite{nustar} consists of two co-aligned telescopes, and the data from both telescopes were fitted simultaneously to a spectral model using XSPEC. For all observations, the source was detected at high significance from the lower end of NuSTAR bandpass ($\approx$3~keV) up to (typically) 40~keV.  The data were reduced using the standard NuSTAR pipeline (NuSTARDAS) with the source extracted from a region $80\arcsec$ in radius around the centroid of the source, and the background subtracted from a similarly sized source-free region. Each pointing was analyzed independently. We found that in all cases, the source was adequately fitted with a power law model.

 Flux variability have been found from one observation to another and the NuSTAR spectrum seems to be steeper when the source is fainter (ranging roughly from $\Gamma=$ 2.5 to 3) - a trend also seen in other HBL objects studied by NuSTAR.

\begin{figure}
\includegraphics[width=85mm]{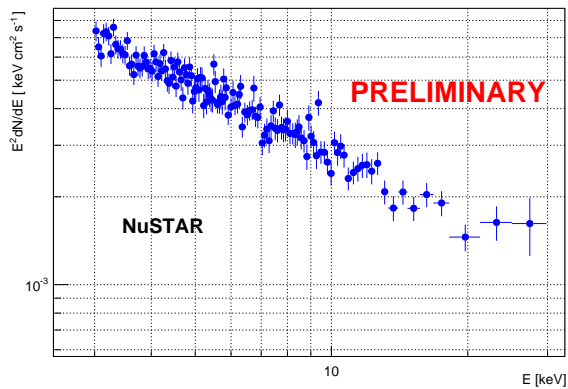}
\caption{Spectral energy distribution measured by NuSTAR.}
\label{NustarSED}
\end{figure}

We also searched for the onset of the inverse Compton component, expected to appear at some point in the hard X-ray/soft gamma-ray spectrum. Here, we searched for statistically significant improvement of the quality of fit with the addition of a hard (assumed $\Gamma_{\rm hard} = 2$) spectral component. There is no clear indication for a presence of such component (primarily because the cosmic X-ray background and instrumental background become more significant at higher energies), and we can limit the flux of this component in the 20--40 keV band to be less than $\sim 10^{-12}$ erg cm$^{-2}$ s$^{-1}$ (Fig. \ref{NustarSED}).

\section{Spectral energy distribution}
\begin{figure}
\includegraphics[width=85mm]{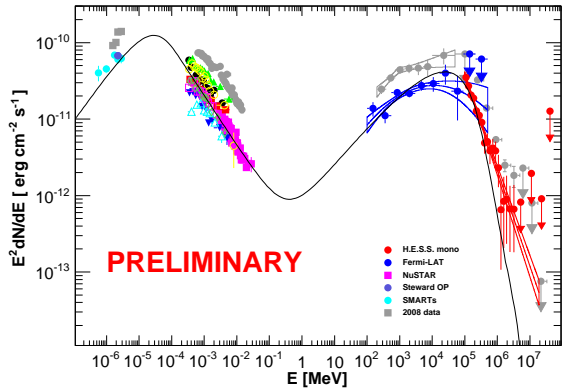}
\caption{Optical to TeV SED of the object. Optical data come from SMARTs and Steward Observatory programs.\textit{Swift} data have been also analysed. The gray points are the archival data from \cite{2008}. The black line is the results of the SSC calculation described in the text.}
\label{SEDMWL}
\end{figure}

The time-averaged SED of the source measured from April to October 2013 is presented on Fig \ref{SEDMWL}. In the optical wavelengths, data from the SMARTs \cite{smart} and Steward Observatory \cite{sop} programs have been used (light blue points). Nine exposures of \textit{Swift}-XRT data have been analysed to extend the X-ray spectrum at lower energies.  The archival data from \cite{2008} are also shown for comparison. The source was at a lower flux state in all wavebands during the 2013 campaign.

A one zone synchrotron self-Compton model \cite{ssc} has been used to reproduce the observed data. The emitting population of electrons $N_e(\gamma)$ is described by a broken power law with an index $p_1=1.8$ below $\gamma=4000$ and $p_2=4.3$ above for a total number of electrons of $4.1\times10^{52}$. The spherical zone of radius $R=7.1\times10^{16}$cm is filled with a constant magnetic field $B=0.05$ Gauss. The emission zone has a bulk Lorentz factor $\delta=20$. The extragalactic background light absorption is taken into account using the model of \cite{fra}.

The ratio of the escape time over the synchrotron cooling time is  0.3, between 0.3 and 3 as recommended by \cite{tav1998}. The particle energy density dominates the magnetic energy with an equipartition factor of $U_{\rm e}/U_{\rm b} \approx 7.3$. The broken power law shape used for the electrons distrubtion can be fitted in a restricted energy range by a curve model since the photon spectra is not strictly a broken power law. A more detailed SSC modelisation will be presented in a forecoming paper.

\section{Conclusion}

The HBL PKS 2155-304 has been observed in 2013 with NuSTAR, \fla\ and H.E.S.S. (CT5) allowing the SED to be measured with an unprecedented precision. The source is found to be at a lower flux state in the $\gamma$-ray regime with respect to the campaign led in 2008. 

No contamination of the X-ray flux by the inverse Compton component has been found in the NuSTAR data. The SED is well described be an one zone SSC model which is self-consistent and with a jet that is particle dominated.

\bigskip % extra skip inserted

\section*{Acknowledgments} \label{Ack}

The support of the Namibian authorities and of the University of Namibia in facilitating the construction and operation of H.E.S.S. is gratefully acknowledged, as is the support by the German Ministry for Education and Research
(BMBF), the Max Planck Society, the French Ministry for Research, the CNRS-IN2P3 and the Astroparticle Interdisciplinary Programme of the CNRS, the U.K. Science and Technology Facilities Council (STFC), the
IPNP of the Charles University, the Polish Ministry of Science and Higher Education, the South African Department of Science and Technology and National Research Foundation, and by the University of Namibia. We appreciate
the excellent work of the technical support staff in Berlin, Durham, Hamburg, Heidelberg, Palaiseau, Paris, Saclay, and in Namibia in the construction and operation of the equipment.

The \textit{Fermi}-LAT Collaboration acknowledges support for LAT development, operation and data analysis from NASA and DOE (United States), CEA/Irfu and IN2P3/CNRS (France), ASI and INFN (Italy), MEXT, KEK, and JAXA (Japan), and the K.A.~Wallenberg Foundation, the Swedish Research Council and the National Space Board (Sweden). Science analysis support in the operations phase from INAF (Italy) and CNES (France) is also gratefully acknowledged.

The work of DS has been supported by the Investissements d'avenir, Labex ENIGMASS.

\end{document}